\documentclass[
apjl,
tighten,
twocolappendix,
twocolumn
]{aastex631}

\usepackage[T1]{fontenc}
\usepackage{CJKutf8}
\usepackage{hyperref}
\usepackage{mathtools}
\usepackage{amsmath} 
\usepackage{physics}
\usepackage{amssymb}
\usepackage{enumerate}
\usepackage{bm}
\usepackage{graphicx}

\shorttitle{
Gamma-ray bursts and kilonova emission from magnetized accretion-induced collapse of white dwarfs
}
\shortauthors{Cheong~\emph{et~al.}}
\graphicspath{{./}{figures/}}

\begin{document}

\title{Gamma-ray bursts and kilonovae from the accretion-induced collapse of white dwarfs}

\author[0000-0003-1449-3363]{Patrick Chi-Kit \surname{Cheong} \begin{CJK*}{UTF8}{bkai}(張志杰)\end{CJK*}}
\email{patrick.cheong@berkeley.edu}
\altaffiliation[]{N3AS Postdoctoral fellow}
\affiliation{Department of Physics \& Astronomy, University of New Hampshire, 9 Library Way, Durham NH 03824, USA}
\affiliation{Center for Nonlinear Studies, Los Alamos National Laboratory, Los Alamos, NM 87545, USA}
\affiliation{Department of Physics, University of California, Berkeley, Berkeley, CA 94720, USA}

\author[0000-0002-9109-2451]{Tetyana \surname{Pitik}}
\email{tetyana.pitik@berkeley.edu}
\altaffiliation[]{N3AS Postdoctoral fellow}
\affiliation{Department of Physics, University of California, Berkeley, Berkeley, CA 94720, USA}
\affiliation{Institute for Gravitation and the Cosmos, The Pennsylvania State University, University Park PA 16802, USA}

\author[0000-0002-7500-7384]{Lu\'is Felipe \surname{Longo Micchi} }
\email{luis.felipe.longo.micchi@uni-jena.de}
\affiliation{Theoretisch-Physikalisches Institut, Friedrich-Schiller-Universit\"at Jena, 07743, Jena, Germany}

\author[0000-0001-6982-1008]{David \surname{Radice}}
\thanks{Alfred P.~Sloan Fellow}
\affiliation{Institute for Gravitation and the Cosmos, The Pennsylvania State University, University Park PA 16802, USA}
\affiliation{Department of Physics, The Pennsylvania State University, University Park PA 16802, USA}
\affiliation{Department of Astronomy \& Astrophysics, The Pennsylvania State University, University Park PA 16802, USA}



\begin{abstract}
We present the first seconds-long 2D general relativistic neutrino magnetohydrodynamic simulations of accretion-induced collapse (AIC) in rapidly rotating, strongly magnetized white dwarfs (WDs), which might originate as remnants of double-WD mergers. This study examines extreme combinations of magnetic fields and rotation rates, motivated both by the need to address the limitations of 2D axisymmetric simulations and to explore the physics of AIC under rare conditions that, while yet to be observationally confirmed, may be consistent with current theoretical models and account for unusual events. Under these assumptions, our results demonstrate that, if realizable, such systems can generate relativistic jets and neutron-rich outflows with properties consistent with long gamma-ray bursts (LGRBs) accompanied by kilonovae, such as GRB 211211A and GRB 230307A. These findings highlight the potential role of AIC in heavy $r$-process element production and offer a framework for understanding rare LGRBs associated with kilonova emission. Longer-duration 3D simulations are needed to fully capture magnetic field amplification, resolve instabilities, and determine the fate of the energy retained by the magnetar at the end of the simulations.
\end{abstract}



\section{\label{sec:intro}Introduction}
The nature of the central engines driving long-duration gamma-ray bursts (LGRBs) accompanied by kilonova emissions, such as GRB 211211A~\citep{2022Natur.612..223R, 2022Natur.612..228T, 2022Natur.612..236M, 2022Natur.612..232Y} and GRB 230307A~\citep{2023GCN.33405....1F, 2023GCN.33407....1D, 2023GCN.33411....1D, 2023GCN.33419....1E, 2023GCN.33429....1B}, remains unresolved.
These events have been suggested to originate from neutron star mergers~\citep{2006Natur.444.1053G, 2022Natur.612..223R, 2022Natur.612..228T, 2022Natur.612..232Y}.
While neutron star mergers are known to power kilonovae via the rapid neutron capture ($r$-process) nucleosynthesis~\citep{1998ApJ...507L..59L, 2010MNRAS.406.2650M, 2016AdAst2016E...8T, 2019LRR....23....1M}, the long-duration GRB is difficult to reconcile with ($\lesssim 1$~s) cooling timescale of the remnant disk formed in these systems \citep[e.g.,][]{2016ARNPS..66...23F}. See however \citet{2023ApJ...958L..33G} for a compact-merger-powered long-GRBs model.
An alternative scenario to explain these events includes the collapse or merger of white dwarfs (WDs; \citealt{1999ApJ...520..650F, 2008MNRAS.385.1455M, 2024ApJ...973L..33C}).
However, such models have not been studied numerically as extensively as compact binary mergers.
 
WDs are compact objects supported by electron degeneracy pressure.
The fate of an accreting WD depends on its mass, composition, and accretion rate~\citep{1986PrPNP..17..249N}.
Low-mass carbon-oxygen WDs ($\lesssim 1.2 \,M_{\odot}$) with high accretion rates are likely to undergo carbon deflagration and explode as Type Ia supernovae, leaving no remnant~\citep{2007MNRAS.380..933Y, 2009ApJ...692..324S, 2009ApJ...705..693S, 2013ApJ...776...97M}.
In contrast, more massive oxygen-neon (O+Ne) WDs ($\gtrsim 1.2 \,M_{\odot}$) with slower accretion rates can result in higher mass WDs.
When the total gravitational mass of a WD exceeds the Chandrasekhar limit of approximately $1.44\,M_{\odot}$~\citep{1931ApJ....74...81C}, gravitational collapse into a neutron star could take place, resulting in AIC event~\citep{1991ApJ...367L..19N}.

The progenitors of AICs are expected to rotate rapidly due to their accretion history~\citep{2003ApJ...583..885P, 2003ApJ...598.1229P, 2003ApJ...595.1094U, 2004ApJ...615..444S, 2018ApJ...869..140K}, and are likely strongly magnetized~\citep{2015SSRv..191..111F, 2020IAUS..357...60K, 2020AdSpR..66.1025F, 2015ApJ...806L...1Z, 2024A&A...691A.179P}.
Such collapses are expected to result in dim and fast evolving transients, and have been suggested as possible sources for next-generation ground-based gravitational-wave observatories~\citep{2008PhRvD..78f4056D, 2023MNRAS.525.6359L}. 
Moreover, they might power gamma-ray bursts~\citep{1998ApJ...494L.163Y, 2008MNRAS.385.1455M, 2009ApJ...696.1871P} and fast radio bursts~\citep{2017ApJ...842...34W, 2019ApJ...886..110M}.
These collapses may further result in the formation of a millisecond pulsar or low-mass X-ray binary~\citep{2015ApJ...800...98A, 2022MNRAS.510.6011W, 2023MNRAS.519.1327A}, and could also be sites of production of rare neutron-rich heavy elements via the $r$-process nucleosynthesis~\citep{1999ApJ...516..892F}.

Despite the significant astrophysical potential of AICs, the current understanding primarily stems from either non-magnetized or Newtonian simulations~\citep{1987ApJ...320..304B, 1992ApJ...391..228W, 1999ApJ...516..892F, 2006ApJ...644.1063D, 2007ApJ...669..585D, 2010PhRvD..81d4012A, 2020ApJ...894..146S, 2023MNRAS.525.6359L, 2023arXiv230617381M}.
To date, no studies have employed general relativistic neutrino magnetohydrodynamics (GR$\nu$MHD) to explore AICs of WDs.
In this work, for the first time, we perform axisymmetric GR$\nu$MHD simulations of rapidly rotating, magnetized WDs with energy-integrated neutrino transport, focusing on jet formation and matter outflows.
The paper is organised as follows.
In section~\ref{sec:methods}, we outline the methods we used in this work.
We present our results in section~\ref{sec:results}, and summarise our findings and conclusions in section~\ref{sec:conclusions}.

\section{\label{sec:methods}Methods}
The primary goal of our study is to explore whether extreme cases of AIC can drive relativistic jets and produce GRBs, which are much rarer than Type Ia SNe. 
Hence, we focus on events where extreme initial conditions may be critical.
Our simulations model a specific subclass of progenitors: highly magnetized, rapidly rotating super-Chandrasekhar WDs formed through the AIC process. 
These WDs differ fundamentally from typical, stable, and observationally constrained WDs. 
The proposed progenitors arise under extreme conditions, which may naturally result from various AIC formation channels, including single-degenerate or double-degenerate WD mergers. 
These scenarios are conducive to substantial magnetic field amplification during accretion or merger phases, potentially driven by dynamo mechanisms linked to the magneto-rotational instability (e.g.,~\cite{2015ApJ...806L...1Z, 2024arXiv240702566P}).
Moreover, the rapid rotation assumed in our models aligns with the expected properties of these super-Chandrasekhar WDs, which may rotate at or beyond the Keplerian limit. 
Consequently, our simulations focus on rapidly rotating, strongly magnetized super-Chandrasekhar WDs as potential progenitors for AICs leading to GRBs.

The initial equilibrium configuration of the rapidly rotating relativistic progenitor used in our simulations are generated with \texttt{RNS} code~\citep{1995ApJ...444..306S}.
We adopt the ``LS220'' equation of state (EOS; \citealt{1991NuPhA.535..331L}). The initial configuration is constructed at a fixed temperature of $0.01~\mathrm{MeV}$ and an electron fraction $Y_e = 0.5$.
The WD has an initial mass of $1.5\,M_{\odot}$, with a central energy density $\epsilon_{c}/c^2 = 1\times10^{10}~\rm{g \cdot cm^{-3}}$~\citep{1991ApJ...367L..19N, 2004A&A...419..623Y, 2005A&A...435..967Y, 2006ApJ...644.1063D, 2007ApJ...669..585D}.
The degree of stellar rotation is characterized by the ratio between the polar and equatorial radii, $a_r$, which reflects the star’s rotational deformation.
Since progenitors of AICs are expected to rotate rapidly due to their accretion history~\citep{2003ApJ...583..885P, 2003ApJ...598.1229P, 2003ApJ...595.1094U, 2004ApJ...615..444S, 2018ApJ...869..140K}, we focus on rapidly rotating WDs with $a_r=0.75$, corresponding to an angular velocity $\Omega \approx 5~{\rm Hz}$, which is $\sim 80\%$ of the Kepler limit.

We endow the initial WD profile with a purely poloidal magnetic field that is roughly uniform inside a sphere of radius $r_{0}=600$~km~\citep{2007ApJ...669..585D}, where the poloidal component is aligned with the axis of rotation.
Such magnetic field is described by the following vector potential \citep[e.g.,][]{2007PASJ...59..771S, 2021MNRAS.508.6033V}: 
\begin{equation}
	\left(A^{\hat{r}}, A^{\hat{\theta}}, A^{\hat{\phi}}\right) = \frac{r_0^3 }{2\left(r^3+r_0^3\right)}\left(0, 0, B_{\rm pol} r \sin\theta \right),
\end{equation}
where $r$ is the distance from the center.
To account for the instabilities that can amplify the magnetic field, and that cannot be resolved due to axisymmetry and limited resolution of our simulations, we consider large initial magnetic fields up to $10^{12}~{\rm G}$.
Specifically, we perform simulations with a purely poloidal magnetic field, using four initial field strengths: $B_{\rm pol} = \left\{ 10^{9}, 10^{10}, 10^{11}, 10^{12}\right\}~{\rm G}$. 
Also, we superimpose a temperature profile $T = T_c \left( \rho / \rho_c \right)^{0.35}$, with $T_c = 5 \times 10^{9} ~{\rm K}$ ($\approx 0.43~{\rm MeV}$) similar to \cite{2006ApJ...644.1063D, 2007ApJ...669..585D} before the dynamical simulations. 

Our models are evolved with the GR$\nu$MHD code \texttt{Gmunu}~\citep{2020CQGra..37n5015C, 2021MNRAS.508.2279C, 2022ApJS..261...22C, 2023ApJS..267...38C, 2024ApJS..272....9N}, which solves GR$\nu$MHD and Einstein field equations in the conformally flat approximation, and uses the energy-integrated two-moment radiation transport scheme of \cite{2022MNRAS.512.1499R} for neutrino transport.
Neutrino rates are provided by coupling \texttt{WeakRates} module in \texttt{WhiskyTHC} code~\citep{2022MNRAS.512.1499R}.
The divergence-free condition of the magnetic field is preserved by using staggered-meshed constrained transport \citep{1988ApJ...332..659E}.

All simulations are axisymmetric and performed in cylindrical coordinates $(R, z)$, with a computational domain extending from $0 \leq R \leq 2000~{\rm km}$ and $0 \leq z \leq 2000~{\rm km}$.
The base resolution is $n_R \times n_z = 128 \times 128$ grid points in $R$ and $z$, with six levels of adaptive mesh refinement applied.
At the center of the star, the finest grid resolution is $\Delta R = \Delta z \approx 488 ~\rm{m}$ when the maximum density is beyond $10^{12}~{\rm g \cdot cm^{-3}}$.

Our simulations adopt the Harten, Lax and van Leer (HLL) approximated Riemann solver~\citep{harten1983upstream}, the 3rd-order reconstruction piecewise parabolic method (PPM)~\citep{1984JCoPh..54..174C} and the IMEXCB3a time integrator~\citep{2015JCoPh.286..172C}. 
The finite temperature equation-of-state ``LS220''~\citep{1991NuPhA.535..331L} is used for the evolution.
To capture the dynamics of the low density gas, we extend the equation-of-state by assuming the gas is a mixture of electrons, ions, and photons.
The details of this extension are presented in the appendix~\ref{sec:low_den}.

Energy-coupled neutrino interactions such as neutrino-electron inelastic scattering, which are important in the collapsing phase, are not included when an energy-integrated neutrino transport is used.
An effective approach to overcome this issue is to apply the parametrised deleptonisation scheme of \cite{2005ApJ...633.1042L} for electron fraction evolutions. 
We switch to the two-moment neutrino transport when the core bounces, and enable the coupling between neutrinos and the fluid, as described in \cite{2024arXiv240716017C}.

We investigate the kilonova emission starting from the ejecta profile extracted from the simulation with the strongest magnetic field, $B_{\rm{pol}}=10^{12}~{\rm G}$.
Unbound material is identified via the Bernoulli criterion and collected at an extraction radius of $r_{\rm{ext}}=1800$~km.
To account for the angular dependence of the ejecta properties and model the long-term evolution, the polar dependence is discretized into 90 angular sections.
Thermodynamic properties such as entropy, temperature, and electron fraction for each angular section are tabulated as a function of the enclosed ejecta mass using a mass-weighted average, procedure adopted also in~\citep{Wu:2021ibi,Magistrelli:2024zmk}.
The time-dependent ejecta profile is mapped into a Lagrangian profile, with the latest shell at the extraction radius and previous layers placed on top.
We solve the spherically symmetric hydrodynamic equations using a ray-by-ray approach, treating each angular section as a 1D problem.
The mass of each shell is scaled by the factor $4\pi/\Delta\Omega$, where $\Delta\Omega = 2\pi \sin\theta \dd \theta$ represents the solid angle of the section.
The position $r(m)$ of each mass shell is computed based on the constraint $m(r) = 4\pi \int_{r_{\rm ext}}^{r} \rho(r) r^2 \dd r$, where $\rho$ is the mass density and $m(r)$ is the enclosed mass.
We neglect non-radial flows of matter and radiation between different sections.
The ejecta profile is evolved using the 1D radiation hydrodynamics code \texttt{SNEC} as implemented in~\citet{Morozova:2015bla} and \citet{Wu:2021ibi}.
Kilonova light curves are calculated using the analytic, time-independent opacity model from~\citet{Wu:2021ibi}.
The results are mapped back into an axisymmetric framework, rescaling the luminosities by $\Delta\Omega/4\pi$, and combining them to calculate the final kilonova emission.

\section{\label{sec:results}Results}

\paragraph{Collapse dynamics} 
The WD collapses due to electron capture and the resulting deleptonization.
The collapse continues until the rest-mass density exceeds nuclear saturation density ($\rho_c \gtrsim 10^{14}~{\rm g \cdot cm^{-3}}$), leading to core bounce and the formation of a proto-neutron star.
Since the collapse and early post-bounce dynamics are consistent with those reported by~\cite{2023MNRAS.525.6359L} across all models, our focus below is mostly on the post-bounce evolution.

\begin{figure*}
	\centering
	\includegraphics[width=\textwidth, angle=0]{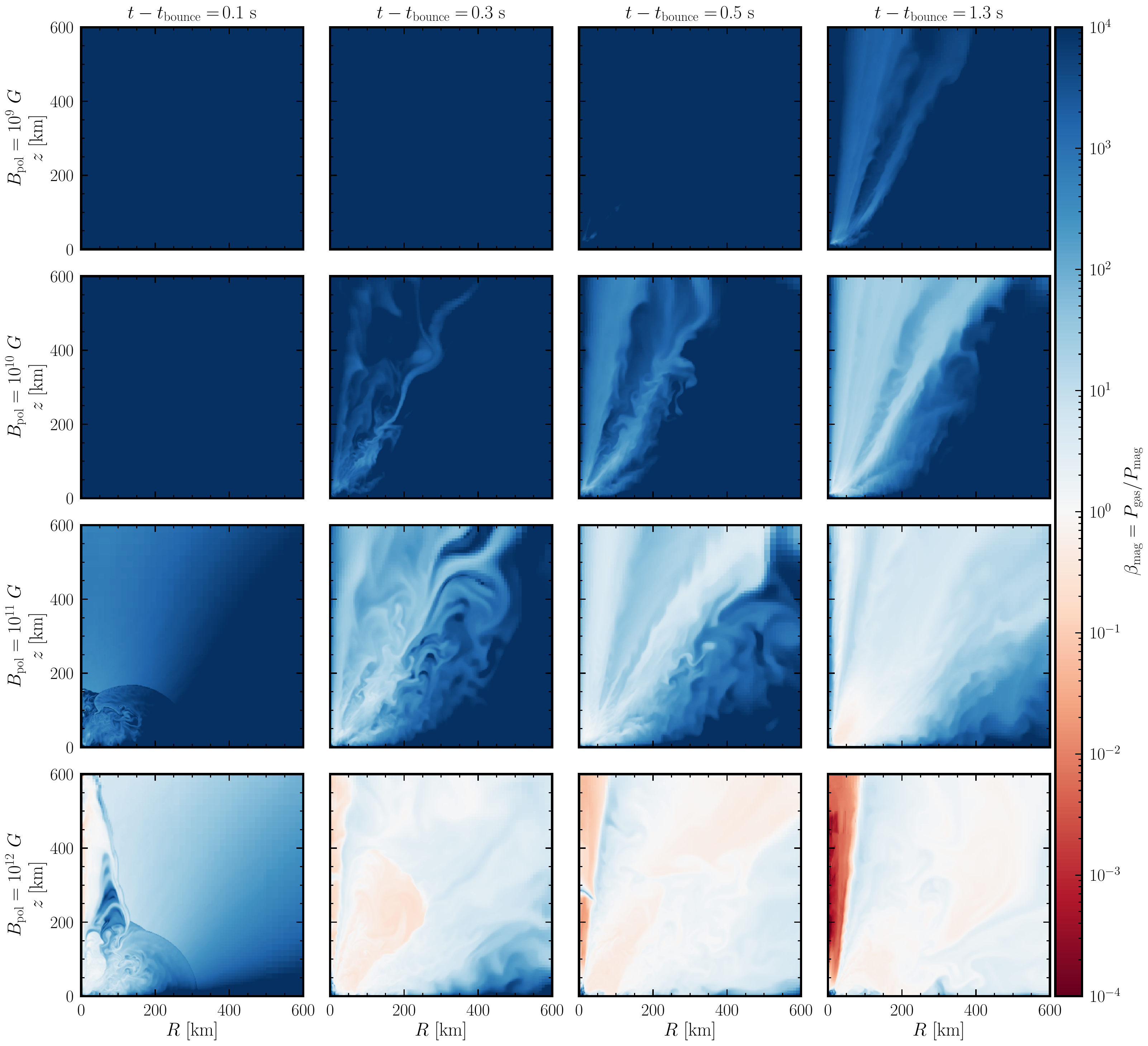}
	\caption{
		Magnetization $\beta_{\rm mag} \coloneqq  P_{\rm gas} / P_{\rm mag}$ profiles (colour map) at different times (\emph{from left to right columns}) of different models (\emph{from top to bottom rows}).
		The blue regions show the matter-dominating regions ($\beta_{\rm mag} > 1$) while the red regions represent the magnetic-dominating ones ($\beta_{\rm mag} < 1$).
		For the $B_{\rm pol} = 10^{12}~{\rm G}$ case, significant parts of the shown regions are magnetic-dominated, and a strongly magnetized jet-like structure is formed at the polar region.
		The corresponding magnetization $\beta_{\rm mag}$ can decrease below $10^{-4}$.
		\label{fig:beta_time_series}
}
\end{figure*}

Although all models experience similar collapse and core bounce dynamics, regardless of the initial magnetic field strength, the post-bounce evolution is significantly different.
During the collapse, the magnetic fields are amplified primarily through magnetic flux conservation, with the maximum field strength increasing by approximately a factor of $\sim 2 \times 10^{3}$ at core bounce.
The amplification of magnetic fields can be further developed after core-bounce due to magnetic winding and magneto-rotational instability~\citep{2006PhRvD..73j4015D, 2015Natur.528..376M}.
Fig.~\ref{fig:beta_time_series} shows the magnetization $\beta_{\rm mag} \coloneqq  P_{\rm gas} / P_{\rm mag}$ profiles at different times of different models.
For the $B_{\rm pol} = 10^{12}~{\rm G}$ case, the magnetization $\beta_{\rm mag}$ is very small even at the early time ($t - t_{\rm bounce} \lesssim 0.1~{\rm s}$), and keep decreasing over the simulation.
However, although the sigma-magnetization parameter 
in the polar regions ($\theta \leq 10^{\circ}$)
is increasing and approaches a value of order unity within the simulation’s timeframe, the material in the polar region is only midly relativistic by the end of the simulation, with $\Gamma_{\infty}\lesssim 2$ (equation~\eqref{eq:Gamma_inf}). 
Therefore, to discuss the potential properties of the associated GRBs, we focus primarily on the jet energy and the proto-neutron star’s spin-down time.

\begin{figure*}
	\centering
	\includegraphics[width=\textwidth, angle=0]{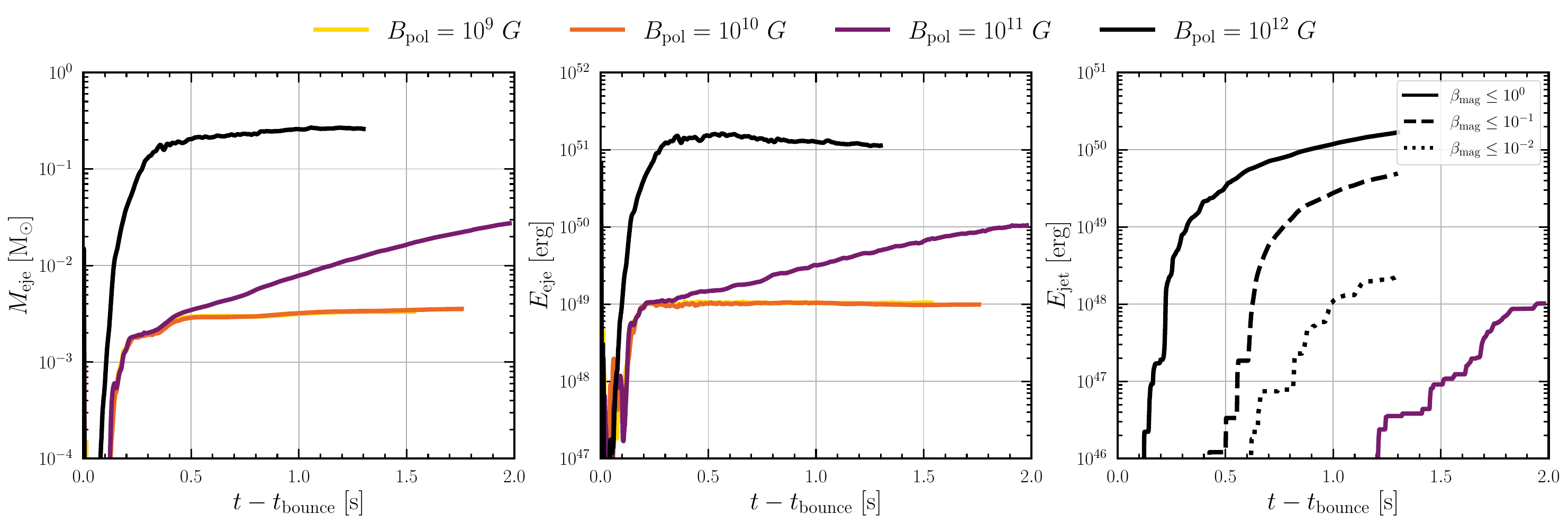}
	\caption{
		The total ejected mass and energy (\emph{left and middle panels}), and the jet energy (\emph{right panel}) of accretion-induced collapse WDs as functions of time.
		Strong magnetically driven winds enhance the total amount of mass and energies.
		The jet energy counts the energy carried away via magnetically-dominated collimated outflow that is leaving along the $z$ direction in the region where $\theta \leq 10^{\circ}$.
		Solid, dashed, and dotted lines show the same diagnostic energy and with extra constraints on the magnetization: $\beta_{\rm mag} \leq 1, 0.1, 0.01$, respectively. 
		\label{fig:eje_mass_and_energy_and_jet_energy}
		}
\end{figure*}

\paragraph{Jets}
To assess the possibility of jet formation, we calculate the energy carried away by the magnetically-dominated ($\beta_{\rm mag} < 1$) collimated ($\theta \leq 10^{\circ}$) outflow along $z$-axis.
The right panel of Fig.~\ref{fig:eje_mass_and_energy_and_jet_energy} shows the estimated jet energy for different models.
For the $B_{\rm pol} = 10^{11}~{\rm G}$ case, the jet energy goes beyond $10^{46}~{\rm erg}$ about $1.25\,\rm{s}$ after bounce, reaching $10^{48}~{\rm erg}$ by the end of the simulation. 
The corresponding isotropic-equivalent energy is: $E_{\rm jet, iso} = E_{\rm jet} / \left[1-\cos\left(\theta_{\rm jet}\right)\right] \approx 6.42 \times 10^{49}~{\rm erg}$, while the luminosity is ${\sim} 9.44\times 10^{51}~{\rm erg \cdot s^{-1}}$.
In contrast, in the $B_{\rm pol} = 10^{12}~{\rm G}$ case, the jet carries more than $10^{45}~{\rm erg}$ energy soon after bounce, and approaches $2 \times 10^{50}~{\rm erg}$ at the end of the simulation.
The corresponding isotropic-equivalent energy and luminosity are ${\sim} 8.03 \times 10^{51}~{\rm erg}$ and ${\sim} 9.18\times 10^{54}~{\rm erg \cdot s^{-1}}$, respectively.
We note that the jet energy is still increasing at the end of the simulations, and we argue that the $B_{\rm pol} = 10^{12}~{\rm G}$ case could eventually release comparable amount of jet energy as in GRB 230307A, whose isotropic-equivalent energy is $\gtrsim 4\times 10^{52}~{\rm erg}$ (see, e.g.,~\cite{2024MNRAS.529L..67D,Rastinejad:2024zuk}).

\begin{figure*}
	\centering
	\includegraphics[width=\textwidth, angle=0]{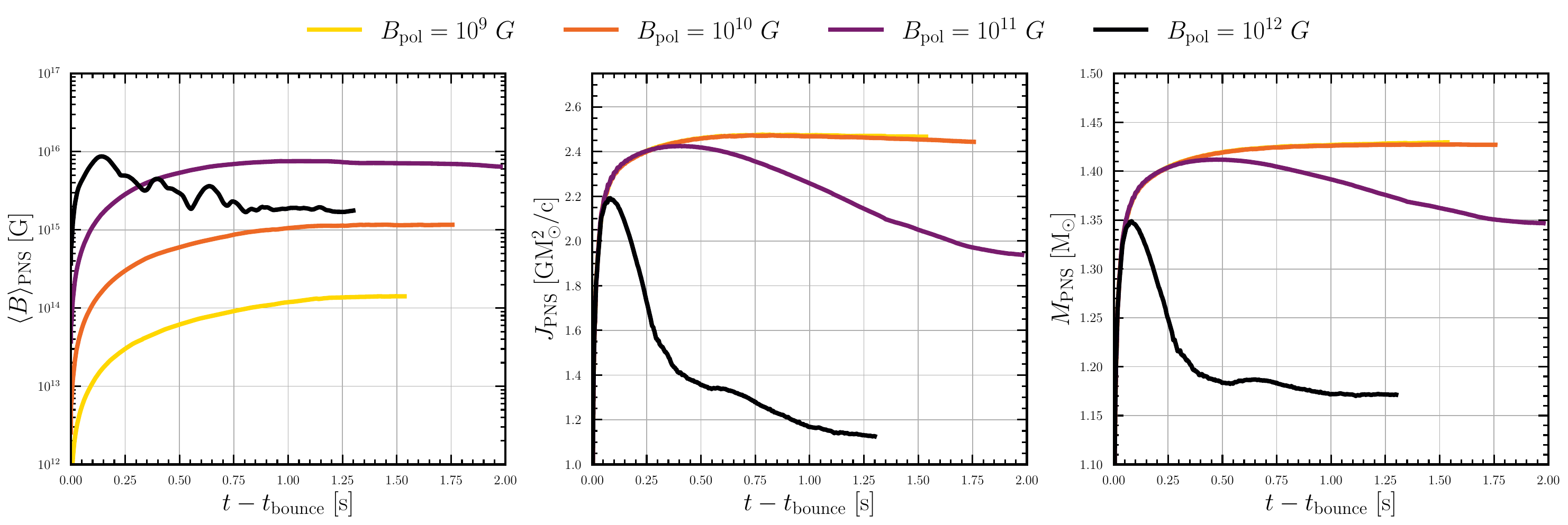}
	\caption{
			Evolution of the rest-mass density averaged magnetic fields $\langle B \rangle _{\rm PNS}$ (\emph{left panel}), angular momentum $J_{\rm PNS}$ (\emph{middle panel}), and rest-mass $M_{\rm PNS}$ (\emph{right panel}) of the proto-neutron stars.
		\label{fig:Bmax_Jpns_Mpns}
		}
\end{figure*}

The jet is powered by the spin-down of the magnetized proto-neutron star.
The duration of the spin-down phase is related to the duration of the gamma-ray burst emission~\citep{2008MNRAS.385.1455M}.
We estimate the spin-up/down time via $\tau \coloneqq J_{\rm PNS}/ \dot{J}_{\rm PNS}$, where $J_{\rm PNS}$ is the angular momentum of the proto-neutron star and $\dot{J}_{\rm PNS}$ is its time derivative.
In all our simulations, the proto-neutron star spins up during the accretion phase, and transitions to a spin-down phase at later times (see e.g. Fig.~\ref{fig:Bmax_Jpns_Mpns}).
The stronger the magnetic fields, the sooner the star starts to spin down.
At the end of the simulations, the spin-down times are $\left\{578, 53, 14, 11\right\}~{\rm s}$ in the $B_{\rm pol} = \left\{ 10^{9}, 10^{10}, 10^{11}, 10^{12}\right\}~{\rm G}$ cases, respectively.
For comparisons, the duration $T_{90}$ (time interval comprising 90\% of the energy emitted in the 50-300 keV band) of the bursts of GRB 211211A and GRB 230307A are about 34~s~\citep{2022Natur.612..236M} and 33~s~\citep{2023ApJ...954L..29D} according to Fermi observatory, respectively.
As the spin-down timescale is mostly affected by the mass ejection, which decrease over time, our values for $\tau$ should be regarded as lower limits an for that we consider our findings for spin-down time of the $B_{\rm pol} = \left\{10^{11}, 10^{12}\right\}~{\rm G}$ cases compatible to the duration of the two GRBs of interest.

\paragraph{Matter outflows}
In addition to the jet, uncollimated outflows are observed in the simulations. 
The left and middle panels of Fig.~\ref{fig:eje_mass_and_energy_and_jet_energy} show a comparison of the total ejected mass and energy across different models. 
By the end of the simulations, the ejected mass and energy in the $B_{\rm pol} = 10^{12}~{\rm G}$ are roughly an order of magnitude higher than in the $B_{\rm pol} = 10^{11}~{\rm G}$ case and two orders of magnitude higher than in the $B_{\rm pol} \leq 10^{10}~{\rm G}$ cases. Our estimated explosion energy is comparable to \cite{2007ApJ...669..585D}.

\begin{figure*}
	\centering
	\includegraphics[width=\textwidth, angle=0]{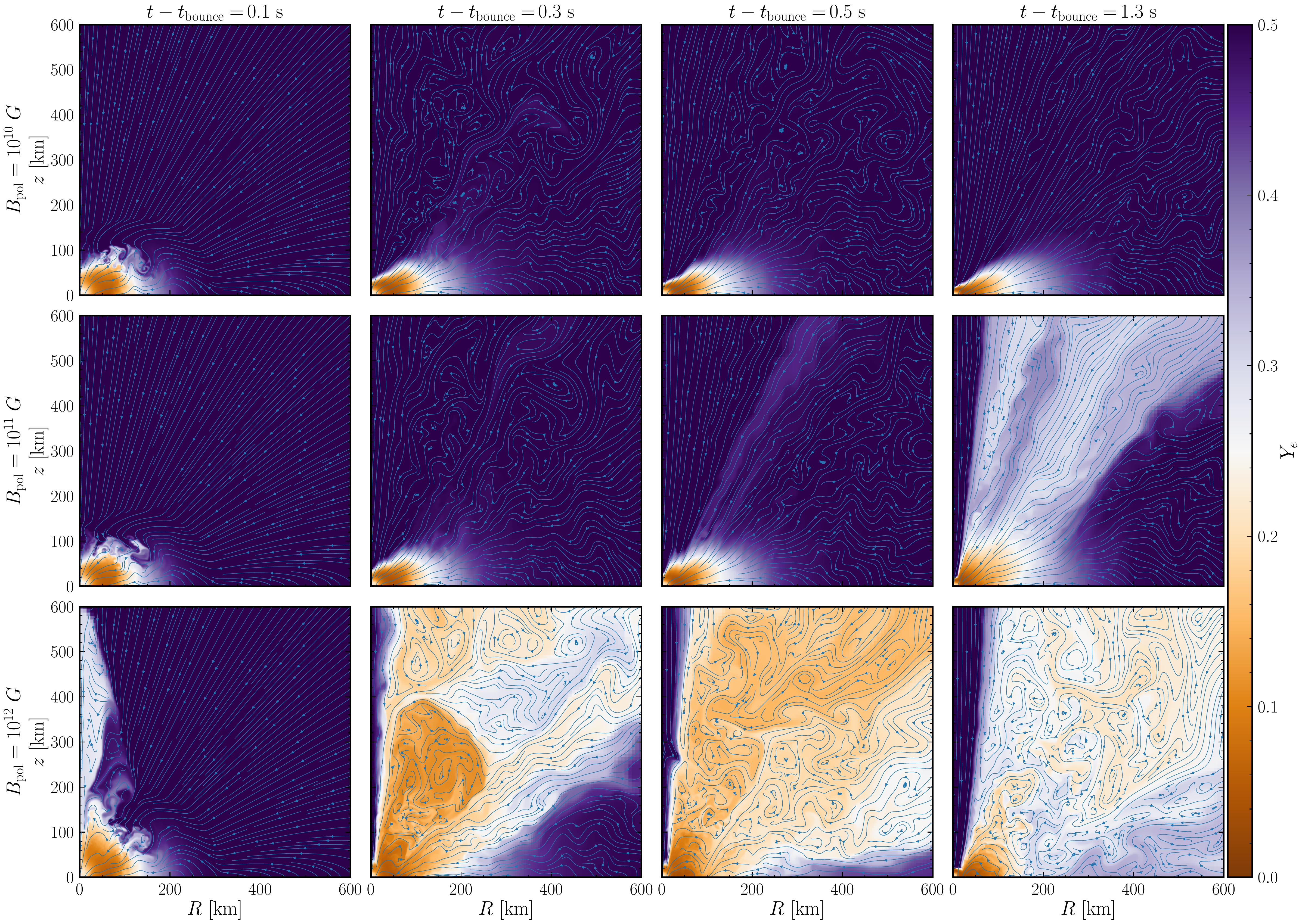}
	\caption{
		Electron fraction $Y_e$ profiles (colour map) and the magnetic field lines (blue streamlines) at different times (\emph{from left to right columns}) for different models (\emph{from top to bottom rows}).
		The $B_{\rm pol} =10^9~{\rm G}$ case is not shown, as it looks identical to the $B_{\rm pol} =10^{10}~{\rm G}$ case.
		Neutron-rich outflow is found when the materials are accelerated out due to strong magnetically driven wind before reaching $\beta$-equilibrium.
		As shown in the figure, when the magnetic field is sufficiently strong (see also Fig.~\ref{fig:beta_time_series}), low $Y_e$ materials can be accelerated and escape the disk region.
		It is clearly visible that the pattern of the electron fraction profile follows the magnetic fields lines. 
		\label{fig:ye_time_series}
		}
\end{figure*}

\begin{figure*}
	\centering
	\includegraphics[width=\textwidth, angle=0]{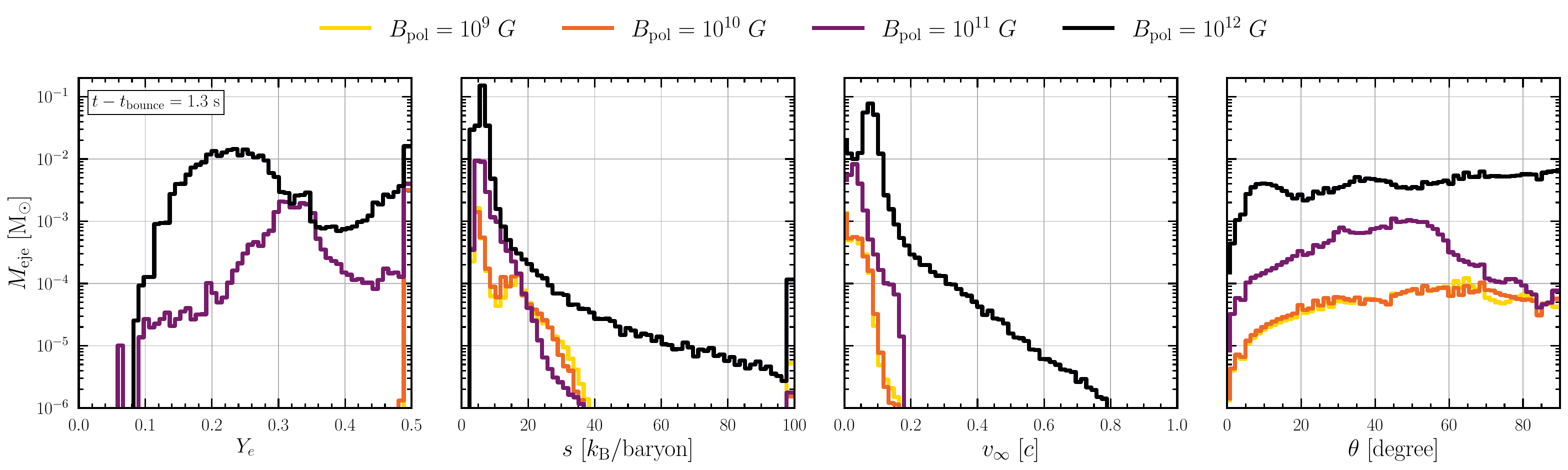}
	\caption{
		1-D histograms of the ejecta of an accretion-induced collapse WD up to 1.3~s after bounce.
		These plots compare the distribution of the ejected mass as functions of electron fraction $Y_e$, entropy per baryon $s$, asymptotic velocity $v_{\infty}$, and angle ($\theta \in \left[ 0^{\circ} , 90^{\circ} \right]$ from pole to equatorial plane) with different initial magnetic fields.
		The stronger the magnetic fields, the more neutron-rich the ejecta are.
		Notably, there is a significant amount of very neutron-rich ejecta (i.e. $0.133~{\rm M_{\odot}}$ for $Y_e \leq 0.25$) in the case of $B_{\rm pol} = 10^{12}~{\rm G}$. 
		\label{fig:eje_histogram}
		}
\end{figure*}

Magnetic fields play a crucial role in shaping matter outflows, as they can accelerate material and affect whether it is ejected before or after reaching weak equilibrium.
Figs.~\ref{fig:ye_time_series} and~\ref{fig:eje_histogram} illustrate the impact of different magnetic field strengths on the ejecta properties.
In the weakly magnetized cases ($B_{\rm pol} \leq 10^{10}~{\rm G}$), the ejecta is predominantly neutron-poor ($Y_e \approx 0.5$) and slow $(v_{\infty} \lesssim 0.2c$).
For $B_{\rm pol} = 10^{11}~\mathrm{G}$, while the early evolution is similar to the $B_{\rm pol} \leq 10^{10}~{\rm G}$ models, the material remains significantly more neutron-rich at later times, with electron fractions ranging from $0.25$ to $0.5$.
In the strongly magnetized case ($B_{\rm pol} = 10^{12}~\mathrm{G}$), the outflow properties differ markedly, with a large amount of extremely neutron-rich material ($\sim 0.133~{\rm M_{\odot}}$ for $Y_e \leq 0.25$$Y_e$), as well as higher entropy and faster-moving ejecta. 
At the end of the simulation, the proto-neutron star continues to rotate rapidly, retaining a significant amount of rotational energy (${\sim} 10^{52}~{\rm erg}$), without collapsing. 
However, it remains unclear how, or if, this energy will be transferred to the ejecta or other channels. 

\paragraph{Kilonova emission} 

Given the significant ejection of material with $Y_e \lesssim 0.25$, the MHD-driven winds of our rotating WD provide an interesting site for $r$-process nucleosynthesis~(\citealt{RevModPhys.29.547, 2015ApJ...815...82L, 2019Natur.569..241S}; see~\citealt{Perego:2021dpw} and \citealt{Cowan:2019pkx} for recent reviews), potentially producing a broad range of nuclei.
Ejecta with low electron fractions are linked to the production of lanthanides and actinides, that can significantly increase the opacity of the ejecta, shifting the emission towards infrared wavelengths~\citep{2013ApJ...775...18B,2013ApJ...775..113T,2015ApJ...815...82L}.
In contrast, ejecta with higher $Y_e$ values ($Y_e \gtrsim 0.3$) produce low-opacity material, which dominates the blue part of the kilonova spectrum, further influenced by neutrino interactions from the central remnant.

\begin{figure*}
	\centering
	\includegraphics[width=0.6\textwidth, angle=0]{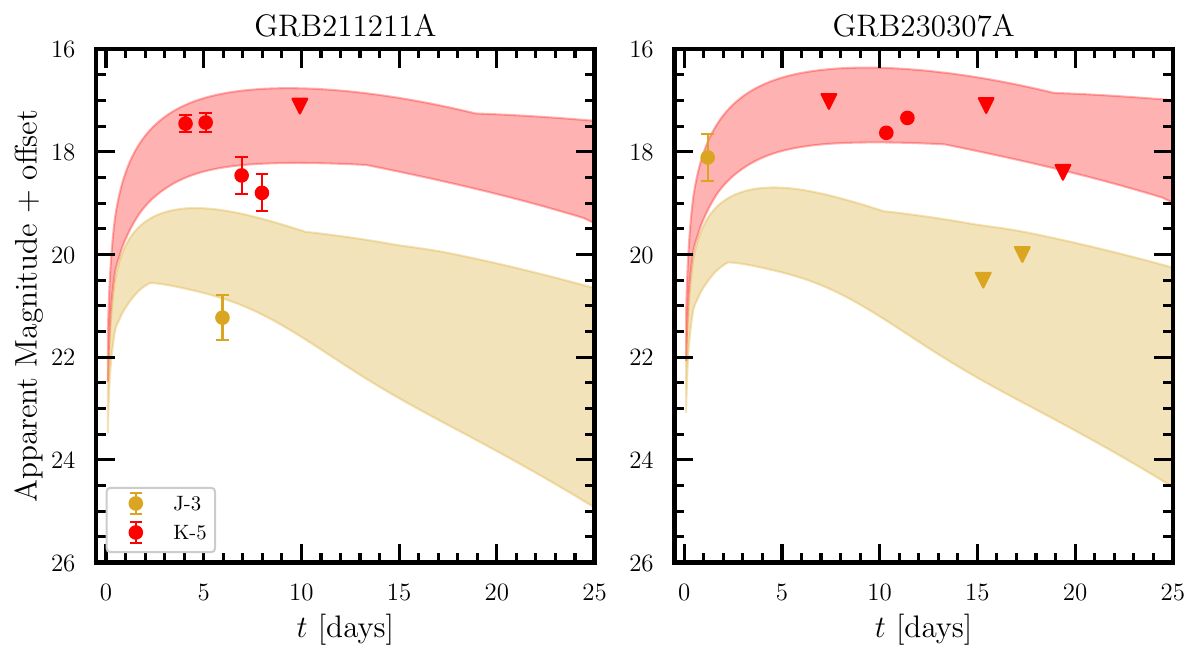}
	\caption{
		GRB 211211A and GRB 230307A data (dots with error bars and upper limits marked by triangles) compared with \texttt{SNEC} AB apparent magnitudes in Gemini J and Ks filters, with offsets applied for readability. 
		The displayed data correspond to the model with $B_{\rm{pol}} = 10^{12}$G (see table~\ref{Table:simulation_results} for ejecta properties).
		The bands account for uncertainties in ejecta velocity, total mass, and maximum opacity.
		We find that our model, with the strongest magnetic field and neutron-rich ejecta, is in good agreement with the observational data.
		Assumed distances are 350 Mpc for GRB 211211A and 291 Mpc for GRB 230307A.
		\label{fig:KN_ligthcurves}
		}
\end{figure*}

In Fig.~\ref{fig:KN_ligthcurves}, we present the results of our kilonova modeling in the AB magnitude system using the J and Ks Gemini filters, where the grey-radiation transport code \texttt{SNEC} is expected to perform better at later times, when the black-body spectrum peaks in the infrared.
The showed bands take into account uncertainties in ejecta opacity, velocity, and total mass.
We vary the maximum opacity between $\kappa = 5~\mathrm{cm^{2} \cdot g^{-1}}$ and $\kappa = 25~\mathrm{cm^{2} \cdot g^{-1}}$, using the $\kappa$ fit as a function of $Y_e$ from~\cite{Wu:2021ibi}.
Given the dynamic nature of the system at the time of data extraction, we account for uncertainties in the radial velocity profile by considering both the instantaneous and asymptotic velocities of the ejecta fluid elements. Since our simulations are time-limited, not all unbound material has yet reached the extraction radius, and additional ejection could occur beyond the simulation’s timeframe. To account for this, we consider the total ejecta mass $M_{\rm{ej}}$ between the amount that had exited the grid by the end of the simulation (${\sim} 1.6 \times 10^{-1}~M_{\odot}$) and a possible upper limit of ${\sim} 3 \times 10^{-1}~M_{\odot}$, reflecting potential later ejections. For reference, the total unbound mass at the end of the simulation is $2.6 \times 10^{-1}M_{\odot}$ (see Table\ref{Table:simulation_results}).
Superimposed over the bands are the afterglow-subtracted observations of GRB 211211A~\citep{2022Natur.612..223R} and GRB 230307A~\citep{2023GCN.33405....1F}, taken from table~4 in~\cite{Rastinejad:2024zuk}.
Overall, our kilonova modeling shows reasonable agreement with the data, following the predicted reddening of the electromagnetic emission over time.
We note that different orientations of the system relative to the observer would further affect the AB magnitudes due to the strong dependence of the $Y_e$ value on the position and polar angle of the ejecta’s fluid elements. However, for simplicity, we do not account for viewing angle effects in this analysis.

\section{\label{sec:conclusions}Discussions and conclusions}
\begin{table}[]
\centering
\hspace{-20mm}
\setlength{\tabcolsep}{2pt}
\makebox[\columnwidth]{
\scalebox{0.85}{
\begin{tabular}{ccccc}
\hline
\hline
Model & $M_{\rm{eje}}$ [$M_\odot$] & $E_{\rm{eje}}$ [erg] & $E_{\rm{jet}}$ [erg] & Spin-down time [s] \\ \hline
$B_{\rm pol} = 10^9~\mathrm{G}$ & $3.3\times 10^{-3}$ & $1.0\times 10^{49}$ & $-$ & $578$ \\ \hline
$B_{\rm pol} = 10^{10}~\mathrm{G}$ & $3.6\times 10^{-3}$ & $9.9\times 10^{48}$ & $-$ & $53$ \\ \hline
$B_{\rm pol} = 10^{11}~\mathrm{G}$ & $2.7\times 10^{-2}$ & $1.05\times 10^{50}$ & $1.02\times 10^{48}$ & $14$ \\ \hline
$B_{\rm pol} = 10^{12}~\mathrm{G}$ & $2.6\times 10^{-1}$ & $1.13\times 10^{51}$ & $1.67\times 10^{50}$ & $11$ \\ \hline
\end{tabular}%
}}
\caption{
	Summary of simulation results.
	Ejecta mass, energy and jet energy are measured at the end of the simulations, corresponding to $1.53\,\rm{s}$, $1.76\,\rm{s}$, $1.98\,\rm{s}$, and $1.3\,\rm{s}$ post-bounce time, respectively.
	}
\label{Table:simulation_results}
\end{table}

In this work, we demonstrated that the accretion-induced collapse of strongly magnetized, rapidly rotating white dwarfs are very promising engines of long gamma-ray bursts associated with kilonova emissions.
We present second-long axisymmetric GR$\nu$MHD simulations of AIC, showing that magnetic fields play a critical role in jet formation and non-relativistic mass ejection, as summarised in table~\ref{Table:simulation_results}.
Moreover, we model the kilonova emission by evolving the initial ejecta profiles via 1D radiation hydrodynamics simulations with realistic opacity and heating rates from radioactive decay.
The AB magnitudes in the Gemini J and Ks bands from our kilonova model, when adjusted for uncertainties in ejecta velocity and total mass, agree well with the observations of GRB 211211A and GRB 230307A.
The evolution of the emission follows the expected reddening over time, as neutron-rich material leads to infrared-dominated light curves.

Our simulations reveal that highly collimated, magnetically dominated outflows are possible in AICs.
In the strongest magnetic field case ($B_{\rm pol} = 10^{12}~\mathrm{G}$), a jet-like funnel forms, with the magnetization parameter ($\beta_{\rm mag}$) dropping below $10^{-4}$.
By the end of the simulation, the jet energy surpasses $10^{50}~\mathrm{erg}$ and continues to increase, with an estimated proto-magnetar spin-down time of $\mathcal{O}(10\,\rm{s})$.
These findings align with the proto-magnetar spin-down model of \cite{2008MNRAS.385.1455M}, supporting the idea that AICs of strongly magnetized WDs can power GRBs.

Magnetically driven winds significantly enhance mass ejection and alter its composition.
In the $B_{\rm{pol}} = 10^{12}\,\rm{G}$ case, much of the ejecta is neutron-rich, with electron fractions ($Y_e$) below 0.25, and some as low as 0.1.
This material has the potential for full $r$-process nucleosynthesis, contributing to heavy element production and kilonova emissions.

While only the $B_{\rm pol} = 10^{12}\,\rm{G}$ case produces a powerful jet energy and neutron-rich ejecta, the $B_{\rm pol} = 10^{11}\rm{G}$ case may also power GRBs and kilonova over longer timescales than those we have simulated.
The mass ejection and jet energy continue to increase by the end of the simulation, and the ejecta becomes progressively more neutron-rich.
With further magnetic field amplification, this case may eventually launch jets and eject considerable amounts of neutron-rich material.

We conclude that AIC of strongly magnetized, rapidly rotating WDs offers a natural explanation for events like GRB 230307A and GRB 211211A.
In the strongest magnetized model, the jet energy and proto-magnetar spin-down timescale align with long GRBs, and the dim explosion combined with neutron-rich ejecta matches the observed light curves.
Thus, AICs are strong candidates for the progenitors of long GRBs with kilonova emissions.

Three-dimensional simulations with higher resolution and longer durations are necessary to further establish the observational implications of our models for LGRBs accompanied by kilonovae. 
To account for the lack of dynamo amplification and other effects in 2D, we have adopted high initial magnetic fields and rapid rotation, as discussed in Sec.~\ref{sec:methods}. 
Future work will address these limitations through longer-term and more detailed 3D simulations, which will enable us to accurately capture magnetic field amplification, jet dynamics, and the role of instabilities such as the kink instability, which are not fully resolved in axisymmetric simulations.

\clearpage

\begin{acknowledgments}
We gratefully thank Cecilia Chirenti for the useful discussions and comments about this manuscript.
P.C.-K.C. and T.P. acknowledge support from NSF Grant PHY-2020275 (Network for Neutrinos, Nuclear Astrophysics, and Symmetries (N3AS)).
LFLM acknowledges funding from the EU Horizon under ERC Consolidator Grant, no.~InspiReM-101043372.
DR acknowledges support from the U.S.~Department of Energy, Office of Science, Division of Nuclear Physics under Award Number(s) DE-SC0021177 and DE-SC0024388, and from the National Science Foundation under Grants No.~PHY-2011725, AST-2108467, PHY-2116686, and PHY-2407681.

The simulations in this work have been performed on the Expanse cluster at San Diego Supercomputer Centre through allocation PHY230104 and PHY230129 from the Advanced Cyberinfrastructure Coordination Ecosystem: Services \& Support (ACCESS) program~\citep{10.1145/3569951.3597559}, which is supported by National Science Foundation grants \#2138259, \#2138286, \#2138307, \#2137603, and \#2138296. 
Additional simulations were performed on NERSC's Perlmutter. 
\texttt{SNEC} simulations have been run on the Pennsylvania State University’s Institute for Computational and Data Sciences’ Roar supercomputer. 
This research used resources of the National Energy Research Scientific Computing Center, a DOE Office of Science User Facility supported by the Office of Science of the U.S.~Department of Energy under Contract No.~DE-AC02-05CH11231.
\end{acknowledgments}

%



\software{
We have modified \texttt{RNS}~\citep{1995ApJ...444..306S} to generate the initial data.
The dynamical simulations of this work were produced by utilising \texttt{Gmunu}~\citep{2020CQGra..37n5015C, 2021MNRAS.508.2279C, 2022ApJS..261...22C, 2023ApJS..267...38C, 2024ApJS..272....9N}.
Neutrino rates are provided by the \texttt{WeakRates} module in \texttt{WhiskyTHC}~\citep{2022MNRAS.512.1499R}.
The ejecta profile is evolved using the 1D radiation hydrodynamics code \texttt{SNEC}~\citep{Morozova:2015bla, Wu:2021ibi}.
The data of the simulations were post-processed and visualised with 
\texttt{yt} \citep{2011ApJS..192....9T},
\texttt{NumPy} \citep{harris2020array}, 
\texttt{pandas} \citep{reback2020pandas, mckinney-proc-scipy-2010},
\texttt{SciPy} \citep{2020SciPy-NMeth} and
\texttt{Matplotlib} \citep{2007CSE.....9...90H, thomas_a_caswell_2023_7697899}.
}


\appendix

\section{\label{sec:low_den}Low density treatment}
The rest-mass density of ejecta can be lower than the lowest allowed rest-mass density of a given nuclear equation-of-state table.
To capture the dynamics of the low density gas, we extend the equation-of-state by assuming the gas as a mixture of electrons, ions, and photons.
In particular, the pressure of the low density gas is given as the sum of ideal-gas and radiation:
\begin{equation}
\begin{aligned}
	P_{\rm low} = & P_{\rm gas} + P_{\rm rad} \\
	   = & \left(\Gamma - 1\right) \rho \beta_v T + {a_{\rm rad}} T^4 /{3},
\end{aligned}
\end{equation}
where $\Gamma$ is the adiabatic index, which is chosen to be $5/3$.
$a_{\rm rad} = 4 \sigma_{\rm SB} / c$ is the radiation constant, and $\sigma_{\rm SB}$ the Stefan-Boltzmann constant.
The inverse specific heat $ \beta_v $ is a function of $ T $ and $ Y_e $ obtained by matching the pressure to the table at the lowest density point (i.e. $P_{\rm low}(\rho_{\min}, T, Y_e) = P_{\rm tab}(\rho_{\min}, T, Y_e)$):
\begin{equation}
	\beta_v(T, Y_e) = \frac{P_{\rm tab}(\rho_{\min}, T, Y_e) - P_{\rm rad}\left(T\right)}{(\Gamma - 1) \rho_{\min} T}.
\end{equation}
The specific internal energy $\varepsilon$ and the sound speed $c_s$ needed for the evolution, are given by
\begin{align}
	\varepsilon &= \varepsilon_{\rm gas} + \varepsilon_{\rm rad} = \beta_v T + {a_{\rm rad}T^4}/{\rho}, \\
	c^2_s &= \left({c^{\rm gas}_s}\right)^2 + \left({c^{\rm rad}_s}\right)^2 
	        =\frac{\Gamma P_{\rm gas}}{\rho + \Gamma \rho \varepsilon_{\rm gas}} + \frac{1}{3}.
\end{align}
The specific entropy $s$ is needed for post-processing, which is given by
\begin{equation}
\begin{aligned}
	s = & {s}_{\rm gas} + {s}_{\rm rad}, \\
    = & \left\{ \ln\left[ \left( \frac{m_{\rm u}}{\rho}\right) \left( \frac{ 2 \pi m_{\rm u} k_{\rm B} T }{h^2}\right)^{{3}/{2}}\right] + \frac{5}{2}\right\} \\
    & + \frac{4}{3}a_{\rm rad} T^3  \left( \frac{m_{\rm u}}{\rho }\right)\left( \frac{1}{k_{\rm B}}\right),
\end{aligned}
\end{equation}
where $k_{\rm B}$ is the Boltzmann constant, $h$ is the Planck constant, and $m_{\rm u}$ is the atomic unit.

In the conserved to primitive conversion, we often need to obtain temperature $T$ from a given specific internal energy $\varepsilon$.
In this case, the temperature can be obtained by solving
\begin{equation}
	f\left(T\right) := 1 - \frac{\beta_v T + {a_{\rm rad}T^4}{/\rho}}{\varepsilon} = 0.
\end{equation}
In practice, we solve this equation via the Newton-Raphson method.

In this work, the floor rest-mass density is set to be 3 orders of magnitude smaller than the lowest available of the equation-of-state table.

\section{\label{sec:diagnostics}Diagnostics}
The matter is identified as unbound when it fulfills the Bernoulli criteria and is outgoing. 
In particular, we locate the unbound matter everywhere in the computational domain by checking
\begin{align}\label{eq:unbound}
	f_{\rm ub} = 
	\begin{cases}
		1, & {\rm if } \; h_{\rm tot} u_t \geq -h_{\min} \;{\rm and} \; v_r > 0 \\
		0, & {\rm otherwise}
	\end{cases},
\end{align}
where $h_{\rm tot} = 1 + \varepsilon + P/\rho + b^2/\rho$ is specific enthalpy with magnetic field contribution, $b^2$ is the contraction of the magnetic fields in the fluid frame, $h_{\min}$ is its minimum allowed values for a given equation-of-state, $u_t = W(- \alpha + \beta_i v^i )$, and $v_r$ is the radial velocity.
This way, $f_{\rm ub}=1$ when the fluid at a point is unbound while equals to zero elsewhere.

To obtain the histograms, we calculate the ejected mass as follows.
The ejected mass per bin is given by
\begin{align}
	\Delta {M}_{\rm eje} = &\int_0^t dt' \oint_{S_{\rm ext}} \hat{v}^i D f_{\rm bin} \dd{A_i} + \int_{V_{\rm ext}} D f_{\rm ub} f_{\rm bin} \dd{V} , \label{eq:matter_flux}
\end{align}
where $\hat{v}^i = \alpha v^i - \beta^i$, and $D=W\rho$ is the conserved rest-mass density.
Here, $f_{\rm bin}=1$ in the target bins, while $f_{\rm bin}=0$ otherwise. 
The total ejected mass can be obtained by $M_{\rm eje} = \sum_{\rm bins} \Delta M_{\rm eje}$.

Similarly, the diagnostic ejecta energy is given by
\begin{align}
	{E}_{\rm eje} = &\int_0^t dt' \oint_{S_{\rm ext}} \hat{v}^i \tau_{\rm eje} \dd{A_i} + \int_{V_{\rm ext}} \tau_{\rm eje} f_{\rm ub} \dd{V}. \label{eq:energy_flux}
\end{align}
Here, $\tau_{\rm eje}$ is the diagnostic ejecta energy density, which is defined as
\begin{align}
	{\tau}_{\rm eje} = \epsilon_{\rm int} + \epsilon_{\rm kin} + \epsilon_{\rm prs} + \epsilon_{\rm EM},
\end{align}
where $\epsilon_{\rm int}$, $\epsilon_{\rm kin}$, $\epsilon_{\rm prs}$ and $\epsilon_{\rm EM}$ are the internal, kinetic, pressure contribution, and electromagnetic energy densities.
They can be obtained by
\begin{align}
	\epsilon_{\rm int} &= \rho W^2 \left( \varepsilon - \varepsilon_0 \right), \\
	\epsilon_{\rm kin} &= \rho W \left( W - 1 \right), \\
	\epsilon_{\rm prs} &= P \left( W^2 - 1 \right), \\
	\epsilon_{\rm EM}  &= B^2 \left( 1 - \frac{1}{2W^2} \right) - \frac{1}{2}\left(B^i v_i\right)^2,
\end{align}
where $\varepsilon$ is the fluid specific internal energy, and $\varepsilon_0$ is a reference zero-point which is obtained by the same rest-mass density and electron fraction but with zero temperature~\citep{2020PhRvD.102l3015B}.

The extraction surfaces $S_{\rm ext}$ of the integration of equations~\eqref{eq:matter_flux} and \eqref{eq:energy_flux} are chosen to be a cylinder with radius $R=1800~{\rm km}$ and $\lvert z \rvert =1800~{\rm km}$, while $V_{\rm ext}$ is the corresponding enclosed region.

The estimated asymptotic Lorentz factor is given by
\begin{equation}
    \label{eq:Gamma_inf}
	\Gamma_{\infty} = - u_t h_{\rm tot},
\end{equation}
where we assume that all the thermal and magnetic energy will be converted into the kinetic energy of the fluid.
The corresponding asymptotic velocity is: $v_{\infty} = \sqrt{1 - 1/\Gamma_{\infty}^2}$.


\bibliography{references}{}

\begin{thebibliography}{}
\expandafter\ifx\csname natexlab\endcsname\relax\def\natexlab#1{#1}\fi
\providecommand{\url}[1]{\href{#1}{#1}}
\providecommand{\dodoi}[1]{doi:~\href{http://doi.org/#1}{\nolinkurl{#1}}}
\providecommand{\doeprint}[1]{\href{http://ascl.net/#1}{\nolinkurl{http://ascl.net/#1}}}
\providecommand{\doarXiv}[1]{\href{https://arxiv.org/abs/#1}{\nolinkurl{https://arxiv.org/abs/#1}}}

\bibitem[{{Abdikamalov} {et~al.}(2010){Abdikamalov}, {Ott}, {Rezzolla},
  {Dessart}, {Dimmelmeier}, {Marek}, \& {Janka}}]{2010PhRvD..81d4012A}
{Abdikamalov}, E.~B., {Ott}, C.~D., {Rezzolla}, L., {et~al.} 2010, \prd, 81,
  044012, \dodoi{10.1103/PhysRevD.81.044012}

\bibitem[{{Ablimit}(2023)}]{2023MNRAS.519.1327A}
{Ablimit}, I. 2023, \mnras, 519, 1327, \dodoi{10.1093/mnras/stac3551}

\bibitem[{{Ablimit} \& {Li}(2015)}]{2015ApJ...800...98A}
{Ablimit}, I., \& {Li}, X.-D. 2015, \apj, 800, 98,
  \dodoi{10.1088/0004-637X/800/2/98}

\bibitem[{{Barnes} \& {Kasen}(2013)}]{2013ApJ...775...18B}
{Barnes}, J., \& {Kasen}, D. 2013, \apj, 775, 18,
  \dodoi{10.1088/0004-637X/775/1/18}

\bibitem[{{Baron} {et~al.}(1987){Baron}, {Cooperstein}, {Kahana}, \&
  {Nomoto}}]{1987ApJ...320..304B}
{Baron}, E., {Cooperstein}, J., {Kahana}, S., \& {Nomoto}, K. 1987, \apj, 320,
  304, \dodoi{10.1086/165542}

\bibitem[{{Betranhandy} \& {O'Connor}(2020)}]{2020PhRvD.102l3015B}
{Betranhandy}, A., \& {O'Connor}, E. 2020, \prd, 102, 123015,
  \dodoi{10.1103/PhysRevD.102.123015}

\bibitem[{Boerner {et~al.}(2023)Boerner, Deems, Furlani, Knuth, \&
  Towns}]{10.1145/3569951.3597559}
Boerner, T.~J., Deems, S., Furlani, T.~R., Knuth, S.~L., \& Towns, J. 2023, in
  Practice and Experience in Advanced Research Computing, PEARC '23 (New York,
  NY, USA: Association for Computing Machinery), 173–176,
  \dodoi{10.1145/3569951.3597559}

\bibitem[{Burbidge {et~al.}(1957)Burbidge, Burbidge, Fowler, \&
  Hoyle}]{RevModPhys.29.547}
Burbidge, E.~M., Burbidge, G.~R., Fowler, W.~A., \& Hoyle, F. 1957, Rev. Mod.
  Phys., 29, 547, \dodoi{10.1103/RevModPhys.29.547}

\bibitem[{{Burrows} {et~al.}(2023){Burrows}, {Gropp}, {Osborne}, {Page},
  {D'Elia}, {Sbarufatti}, {D'Ai}, {Dichiara}, {Evans}, \& {Swift-XRT
  Team}}]{2023GCN.33429....1B}
{Burrows}, D.~N., {Gropp}, J.~D., {Osborne}, J.~P., {et~al.} 2023, GRB
  Coordinates Network, 33429, 1

\bibitem[{Caswell {et~al.}(2023)Caswell, Lee, de~Andrade, Droettboom, Hoffmann,
  Klymak, Hunter, Firing, Stansby, Varoquaux, Nielsen, Root, May, Gustafsson,
  Elson, Seppänen, Lee, Dale, hannah, McDougall, Straw, Hobson, Sunden, Lucas,
  Gohlke, Vincent, Yu, Ma, Silvester, \& Moad}]{thomas_a_caswell_2023_7697899}
Caswell, T.~A., Lee, A., de~Andrade, E.~S., {et~al.} 2023,
  matplotlib/matplotlib: REL: v3.7.1, v3.7.1,  Zenodo,
  \dodoi{10.5281/zenodo.7697899}

\bibitem[{{Cavaglieri} \& {Bewley}(2015)}]{2015JCoPh.286..172C}
{Cavaglieri}, D., \& {Bewley}, T. 2015, Journal of Computational Physics, 286,
  172, \dodoi{10.1016/j.jcp.2015.01.031}

\bibitem[{{Chandrasekhar}(1931)}]{1931ApJ....74...81C}
{Chandrasekhar}, S. 1931, \apj, 74, 81, \dodoi{10.1086/143324}

\bibitem[{{Chen} {et~al.}(2024){Chen}, {Shen}, {Tan}, {Wang}, {Xiong}, {Chen},
  \& {Zhang}}]{2024ApJ...973L..33C}
{Chen}, J., {Shen}, R.-F., {Tan}, W.-J., {et~al.} 2024, \apjl, 973, L33,
  \dodoi{10.3847/2041-8213/ad7737}

\bibitem[{{Cheong} {et~al.}(2024){Cheong}, {Foucart}, {Duez}, {Offermans},
  {Muhammed}, \& {Chawhan}}]{2024arXiv240716017C}
{Cheong}, C.-K.~P., {Foucart}, F., {Duez}, M.~D., {et~al.} 2024, arXiv
  e-prints, arXiv:2407.16017, \dodoi{10.48550/arXiv.2407.16017}

\bibitem[{{Cheong} {et~al.}(2021){Cheong}, {Lam}, {Ng}, \&
  {Li}}]{2021MNRAS.508.2279C}
{Cheong}, P. C.-K., {Lam}, A. T.-L., {Ng}, H. H.-Y., \& {Li}, T. G.~F. 2021,
  \mnras, 508, 2279, \dodoi{10.1093/mnras/stab2606}

\bibitem[{{Cheong} {et~al.}(2020){Cheong}, {Lin}, \&
  {Li}}]{2020CQGra..37n5015C}
{Cheong}, P. C.-K., {Lin}, L.-M., \& {Li}, T. G.~F. 2020, Classical and Quantum
  Gravity, 37, 145015, \dodoi{10.1088/1361-6382/ab8e9c}

\bibitem[{{Cheong} {et~al.}(2023){Cheong}, {Ng}, {Lam}, \&
  {Li}}]{2023ApJS..267...38C}
{Cheong}, P. C.-K., {Ng}, H. H.-Y., {Lam}, A. T.-L., \& {Li}, T. G.~F. 2023,
  \apjs, 267, 38, \dodoi{10.3847/1538-4365/acd931}

\bibitem[{{Cheong} {et~al.}(2022){Cheong}, {Pong}, {Yip}, \&
  {Li}}]{2022ApJS..261...22C}
{Cheong}, P. C.-K., {Pong}, D. Y.~T., {Yip}, A. K.~L., \& {Li}, T. G.~F. 2022,
  \apjs, 261, 22, \dodoi{10.3847/1538-4365/ac6cec}

\bibitem[{{Colella} \& {Woodward}(1984)}]{1984JCoPh..54..174C}
{Colella}, P., \& {Woodward}, P.~R. 1984, Journal of Computational Physics, 54,
  174, \dodoi{10.1016/0021-9991(84)90143-8}

\bibitem[{Cowan {et~al.}(2021)Cowan, Sneden, Lawler, Aprahamian, Wiescher,
  Langanke, Mart\'\i{}nez-Pinedo, \& Thielemann}]{Cowan:2019pkx}
Cowan, J.~J., Sneden, C., Lawler, J.~E., {et~al.} 2021, Rev. Mod. Phys., 93,
  15002, \dodoi{10.1103/RevModPhys.93.015002}

\bibitem[{{Dalessi} \& {Fermi GBM Team}(2023)}]{2023GCN.33407....1D}
{Dalessi}, S., \& {Fermi GBM Team}. 2023, GRB Coordinates Network, 33407, 1

\bibitem[{{Dalessi} {et~al.}(2023){Dalessi}, {Roberts}, {Meegan}, \& {Fermi GBM
  Team}}]{2023GCN.33411....1D}
{Dalessi}, S., {Roberts}, O.~J., {Meegan}, C., \& {Fermi GBM Team}. 2023, GRB
  Coordinates Network, 33411, 1

\bibitem[{{Dessart} {et~al.}(2007){Dessart}, {Burrows}, {Livne}, \&
  {Ott}}]{2007ApJ...669..585D}
{Dessart}, L., {Burrows}, A., {Livne}, E., \& {Ott}, C.~D. 2007, \apj, 669,
  585, \dodoi{10.1086/521701}

\bibitem[{{Dessart} {et~al.}(2006){Dessart}, {Burrows}, {Ott}, {Livne}, {Yoon},
  \& {Langer}}]{2006ApJ...644.1063D}
{Dessart}, L., {Burrows}, A., {Ott}, C.~D., {et~al.} 2006, \apj, 644, 1063,
  \dodoi{10.1086/503626}

\bibitem[{{Dichiara} {et~al.}(2023){Dichiara}, {Tsang}, {Troja}, {Neill},
  {Norris}, \& {Yang}}]{2023ApJ...954L..29D}
{Dichiara}, S., {Tsang}, D., {Troja}, E., {et~al.} 2023, \apjl, 954, L29,
  \dodoi{10.3847/2041-8213/acf21d}

\bibitem[{{Dimmelmeier} {et~al.}(2008){Dimmelmeier}, {Ott}, {Marek}, \&
  {Janka}}]{2008PhRvD..78f4056D}
{Dimmelmeier}, H., {Ott}, C.~D., {Marek}, A., \& {Janka}, H.~T. 2008, \prd, 78,
  064056, \dodoi{10.1103/PhysRevD.78.064056}

\bibitem[{{Du} {et~al.}(2024){Du}, {L{\"u}}, {Liu}, \&
  {Liang}}]{2024MNRAS.529L..67D}
{Du}, Z.-W., {L{\"u}}, H., {Liu}, X., \& {Liang}, E. 2024, \mnras, 529, L67,
  \dodoi{10.1093/mnrasl/slad203}

\bibitem[{{Duez} {et~al.}(2006){Duez}, {Liu}, {Shapiro}, {Shibata}, \&
  {Stephens}}]{2006PhRvD..73j4015D}
{Duez}, M.~D., {Liu}, Y.~T., {Shapiro}, S.~L., {Shibata}, M., \& {Stephens},
  B.~C. 2006, \prd, 73, 104015, \dodoi{10.1103/PhysRevD.73.104015}

\bibitem[{{Evans} \& {Hawley}(1988)}]{1988ApJ...332..659E}
{Evans}, C.~R., \& {Hawley}, J.~F. 1988, \apj, 332, 659, \dodoi{10.1086/166684}

\bibitem[{{Evans} \& {Swift Team}(2023)}]{2023GCN.33419....1E}
{Evans}, P.~A., \& {Swift Team}. 2023, GRB Coordinates Network, 33419, 1

\bibitem[{{Fermi GBM Team}(2023)}]{2023GCN.33405....1F}
{Fermi GBM Team}. 2023, GRB Coordinates Network, 33405, 1

\bibitem[{{Fern{\'a}ndez} \& {Metzger}(2016)}]{2016ARNPS..66...23F}
{Fern{\'a}ndez}, R., \& {Metzger}, B.~D. 2016, Annual Review of Nuclear and
  Particle Science, 66, 23, \dodoi{10.1146/annurev-nucl-102115-044819}

\bibitem[{{Ferrario} {et~al.}(2015){Ferrario}, {de Martino}, \&
  {G{\"a}nsicke}}]{2015SSRv..191..111F}
{Ferrario}, L., {de Martino}, D., \& {G{\"a}nsicke}, B.~T. 2015, \ssr, 191,
  111, \dodoi{10.1007/s11214-015-0152-0}

\bibitem[{{Ferrario} {et~al.}(2020){Ferrario}, {Wickramasinghe}, \&
  {Kawka}}]{2020AdSpR..66.1025F}
{Ferrario}, L., {Wickramasinghe}, D., \& {Kawka}, A. 2020, Advances in Space
  Research, 66, 1025, \dodoi{10.1016/j.asr.2019.11.012}

\bibitem[{{Fryer} {et~al.}(1999{\natexlab{a}}){Fryer}, {Benz}, {Herant}, \&
  {Colgate}}]{1999ApJ...516..892F}
{Fryer}, C., {Benz}, W., {Herant}, M., \& {Colgate}, S.~A. 1999{\natexlab{a}},
  \apj, 516, 892, \dodoi{10.1086/307119}

\bibitem[{{Fryer} {et~al.}(1999{\natexlab{b}}){Fryer}, {Woosley}, {Herant}, \&
  {Davies}}]{1999ApJ...520..650F}
{Fryer}, C.~L., {Woosley}, S.~E., {Herant}, M., \& {Davies}, M.~B.
  1999{\natexlab{b}}, \apj, 520, 650, \dodoi{10.1086/307467}

\bibitem[{{Gal-Yam} {et~al.}(2006){Gal-Yam}, {Fox}, {Price}, {Ofek}, {Davis},
  {Leonard}, {Soderberg}, {Schmidt}, {Lewis}, {Peterson}, {Kulkarni}, {Berger},
  {Cenko}, {Sari}, {Sharon}, {Frail}, {Moon}, {Brown}, {Cucchiara}, {Harrison},
  {Piran}, {Persson}, {McCarthy}, {Penprase}, {Chevalier}, \&
  {MacFadyen}}]{2006Natur.444.1053G}
{Gal-Yam}, A., {Fox}, D.~B., {Price}, P.~A., {et~al.} 2006, \nat, 444, 1053,
  \dodoi{10.1038/nature05373}

\bibitem[{{Gottlieb} {et~al.}(2023){Gottlieb}, {Metzger}, {Quataert}, {Issa},
  {Martineau}, {Foucart}, {Duez}, {Kidder}, {Pfeiffer}, \&
  {Scheel}}]{2023ApJ...958L..33G}
{Gottlieb}, O., {Metzger}, B.~D., {Quataert}, E., {et~al.} 2023, \apjl, 958,
  L33, \dodoi{10.3847/2041-8213/ad096e}

\bibitem[{Harris {et~al.}(2020)Harris, Millman, van~der Walt, Gommers,
  Virtanen, Cournapeau, Wieser, Taylor, Berg, Smith, Kern, Picus, Hoyer, van
  Kerkwijk, Brett, Haldane, del R{\'{i}}o, Wiebe, Peterson,
  G{\'{e}}rard-Marchant, Sheppard, Reddy, Weckesser, Abbasi, Gohlke, \&
  Oliphant}]{harris2020array}
Harris, C.~R., Millman, K.~J., van~der Walt, S.~J., {et~al.} 2020, Nature, 585,
  357, \dodoi{10.1038/s41586-020-2649-2}

\bibitem[{Harten {et~al.}(1983)Harten, Lax, \& Leer}]{harten1983upstream}
Harten, A., Lax, P., \& Leer, B. 1983, SIAM Review, 25, 35,
  \dodoi{10.1137/1025002}

\bibitem[{{Hunter}(2007)}]{2007CSE.....9...90H}
{Hunter}, J.~D. 2007, Computing in Science and Engineering, 9, 90,
  \dodoi{10.1109/MCSE.2007.55}

\bibitem[{{Kashyap} {et~al.}(2018){Kashyap}, {Haque}, {Lor{\'e}n-Aguilar},
  {Garc{\'\i}a-Berro}, \& {Fisher}}]{2018ApJ...869..140K}
{Kashyap}, R., {Haque}, T., {Lor{\'e}n-Aguilar}, P., {Garc{\'\i}a-Berro}, E.,
  \& {Fisher}, R. 2018, \apj, 869, 140, \dodoi{10.3847/1538-4357/aaedb7}

\bibitem[{{Kawka}(2020)}]{2020IAUS..357...60K}
{Kawka}, A. 2020, IAU Symposium, 357, 60, \dodoi{10.1017/S1743921320000745}

\bibitem[{{Lattimer} \& {Swesty}(1991)}]{1991NuPhA.535..331L}
{Lattimer}, J.~M., \& {Swesty}, D.~F. 1991, \nphysa, 535, 331,
  \dodoi{10.1016/0375-9474(91)90452-C}

\bibitem[{{Li} \& {Paczy{\'n}ski}(1998)}]{1998ApJ...507L..59L}
{Li}, L.-X., \& {Paczy{\'n}ski}, B. 1998, \apjl, 507, L59,
  \dodoi{10.1086/311680}

\bibitem[{{Liebend{\"o}rfer}(2005)}]{2005ApJ...633.1042L}
{Liebend{\"o}rfer}, M. 2005, \apj, 633, 1042, \dodoi{10.1086/466517}

\bibitem[{{Lippuner} \& {Roberts}(2015)}]{2015ApJ...815...82L}
{Lippuner}, J., \& {Roberts}, L.~F. 2015, \apj, 815, 82,
  \dodoi{10.1088/0004-637X/815/2/82}

\bibitem[{{Longo Micchi} {et~al.}(2023){Longo Micchi}, {Radice}, \&
  {Chirenti}}]{2023MNRAS.525.6359L}
{Longo Micchi}, L.~F., {Radice}, D., \& {Chirenti}, C. 2023, \mnras, 525, 6359,
  \dodoi{10.1093/mnras/stad2420}

\bibitem[{Magistrelli {et~al.}(2024)Magistrelli, Bernuzzi, Perego, \&
  Radice}]{Magistrelli:2024zmk}
Magistrelli, F., Bernuzzi, S., Perego, A., \& Radice, D. 2024, Astrophys. J.
  Lett., 974, L5, \dodoi{10.3847/2041-8213/ad74e0}

\bibitem[{{Margalit} {et~al.}(2019){Margalit}, {Berger}, \&
  {Metzger}}]{2019ApJ...886..110M}
{Margalit}, B., {Berger}, E., \& {Metzger}, B.~D. 2019, \apj, 886, 110,
  \dodoi{10.3847/1538-4357/ab4c31}

\bibitem[{{Mei} {et~al.}(2022){Mei}, {Banerjee}, {Oganesyan}, {Salafia},
  {Giarratana}, {Branchesi}, {D'Avanzo}, {Campana}, {Ghirlanda}, {Ronchini},
  {Shukla}, \& {Tiwari}}]{2022Natur.612..236M}
{Mei}, A., {Banerjee}, B., {Oganesyan}, G., {et~al.} 2022, \nat, 612, 236,
  \dodoi{10.1038/s41586-022-05404-7}

\bibitem[{{Metzger}(2019)}]{2019LRR....23....1M}
{Metzger}, B.~D. 2019, Living Reviews in Relativity, 23, 1,
  \dodoi{10.1007/s41114-019-0024-0}

\bibitem[{{Metzger} {et~al.}(2008){Metzger}, {Quataert}, \&
  {Thompson}}]{2008MNRAS.385.1455M}
{Metzger}, B.~D., {Quataert}, E., \& {Thompson}, T.~A. 2008, \mnras, 385, 1455,
  \dodoi{10.1111/j.1365-2966.2008.12923.x}

\bibitem[{{Metzger} {et~al.}(2010){Metzger}, {Mart{\'\i}nez-Pinedo}, {Darbha},
  {Quataert}, {Arcones}, {Kasen}, {Thomas}, {Nugent}, {Panov}, \&
  {Zinner}}]{2010MNRAS.406.2650M}
{Metzger}, B.~D., {Mart{\'\i}nez-Pinedo}, G., {Darbha}, S., {et~al.} 2010,
  \mnras, 406, 2650, \dodoi{10.1111/j.1365-2966.2010.16864.x}

\bibitem[{{Moore} {et~al.}(2013){Moore}, {Townsley}, \&
  {Bildsten}}]{2013ApJ...776...97M}
{Moore}, K., {Townsley}, D.~M., \& {Bildsten}, L. 2013, \apj, 776, 97,
  \dodoi{10.1088/0004-637X/776/2/97}

\bibitem[{{Mori} {et~al.}(2023){Mori}, {Sawada}, {Suwa}, {Tanikawa},
  {Kashiyama}, \& {Murase}}]{2023arXiv230617381M}
{Mori}, M., {Sawada}, R., {Suwa}, Y., {et~al.} 2023, arXiv e-prints,
  arXiv:2306.17381, \dodoi{10.48550/arXiv.2306.17381}

\bibitem[{Morozova {et~al.}(2015)Morozova, Piro, Renzo, Ott, Clausen, Couch,
  Ellis, \& Roberts}]{Morozova:2015bla}
Morozova, V., Piro, A.~L., Renzo, M., {et~al.} 2015, Astrophys. J., 814, 63,
  \dodoi{10.1088/0004-637X/814/1/63}

\bibitem[{{M{\"o}sta} {et~al.}(2015){M{\"o}sta}, {Ott}, {Radice}, {Roberts},
  {Schnetter}, \& {Haas}}]{2015Natur.528..376M}
{M{\"o}sta}, P., {Ott}, C.~D., {Radice}, D., {et~al.} 2015, \nat, 528, 376,
  \dodoi{10.1038/nature15755}

\bibitem[{{Ng} {et~al.}(2024){Ng}, {Cheong}, {Lam}, \&
  {Li}}]{2024ApJS..272....9N}
{Ng}, H. H.-Y., {Cheong}, P. C.-K., {Lam}, A. T.-L., \& {Li}, T. G.~F. 2024,
  \apjs, 272, 9, \dodoi{10.3847/1538-4365/ad2fbd}

\bibitem[{{Nomoto}(1986)}]{1986PrPNP..17..249N}
{Nomoto}, K. 1986, Progress in Particle and Nuclear Physics, 17, 249,
  \dodoi{10.1016/0146-6410(86)90020-7}

\bibitem[{{Nomoto} \& {Kondo}(1991)}]{1991ApJ...367L..19N}
{Nomoto}, K., \& {Kondo}, Y. 1991, \apjl, 367, L19, \dodoi{10.1086/185922}

\bibitem[{{Pakmor} {et~al.}(2024{\natexlab{a}}){Pakmor}, {Pelisoli}, {Justham},
  {Rajamuthukumar}, {R{\"o}pke}, {Schneider}, {de Mink}, {Ohlmann},
  {Podsiadlowski}, {Mor{\'a}n-Fraile}, {Vetter}, \&
  {Andrassy}}]{2024A&A...691A.179P}
{Pakmor}, R., {Pelisoli}, I., {Justham}, S., {et~al.} 2024{\natexlab{a}}, \aap,
  691, A179, \dodoi{10.1051/0004-6361/202451352}

\bibitem[{{Pakmor} {et~al.}(2024{\natexlab{b}}){Pakmor}, {Pelisoli}, {Justham},
  {Rajamuthukumar}, {R{\"o}pke}, {Schneider}, {de Mink}, {Ohlmann},
  {Podsiadlowski}, {Moran Fraile}, {Vetter}, \&
  {Andrassy}}]{2024arXiv240702566P}
---. 2024{\natexlab{b}}, arXiv e-prints, arXiv:2407.02566,
  \dodoi{10.48550/arXiv.2407.02566}

\bibitem[{pandas~development team(2020)}]{reback2020pandas}
pandas~development team, T. 2020, pandas-dev/pandas: Pandas, latest,  Zenodo,
  \dodoi{10.5281/zenodo.3509134}

\bibitem[{Perego {et~al.}(2022)Perego, Thielemann, \&
  Cescutti}]{Perego:2021dpw}
Perego, A., Thielemann, F.-K., \& Cescutti, G. 2022, {r-Process Nucleosynthesis
  from Compact Binary Mergers}, \dodoi{10.1007/978-981-16-4306-4_13}

\bibitem[{{Perley} {et~al.}(2009){Perley}, {Metzger}, {Granot}, {Butler},
  {Sakamoto}, {Ramirez-Ruiz}, {Levan}, {Bloom}, {Miller}, {Bunker}, {Chen},
  {Filippenko}, {Gehrels}, {Glazebrook}, {Hall}, {Hurley}, {Kocevski}, {Li},
  {Lopez}, {Norris}, {Piro}, {Poznanski}, {Prochaska}, {Quataert}, \&
  {Tanvir}}]{2009ApJ...696.1871P}
{Perley}, D.~A., {Metzger}, B.~D., {Granot}, J., {et~al.} 2009, \apj, 696,
  1871, \dodoi{10.1088/0004-637X/696/2/1871}

\bibitem[{{Piersanti} {et~al.}(2003{\natexlab{a}}){Piersanti}, {Gagliardi},
  {Iben}, \& {Tornamb{\'e}}}]{2003ApJ...583..885P}
{Piersanti}, L., {Gagliardi}, S., {Iben}, Icko, J., \& {Tornamb{\'e}}, A.
  2003{\natexlab{a}}, \apj, 583, 885, \dodoi{10.1086/345444}

\bibitem[{{Piersanti} {et~al.}(2003{\natexlab{b}}){Piersanti}, {Gagliardi},
  {Iben}, \& {Tornamb{\'e}}}]{2003ApJ...598.1229P}
---. 2003{\natexlab{b}}, \apj, 598, 1229, \dodoi{10.1086/378952}

\bibitem[{{Radice} {et~al.}(2022){Radice}, {Bernuzzi}, {Perego}, \&
  {Haas}}]{2022MNRAS.512.1499R}
{Radice}, D., {Bernuzzi}, S., {Perego}, A., \& {Haas}, R. 2022, \mnras, 512,
  1499, \dodoi{10.1093/mnras/stac589}

\bibitem[{Rastinejad {et~al.}(2024)Rastinejad, Fong, Kilpatrick, Nicholl, \&
  Metzger}]{Rastinejad:2024zuk}
Rastinejad, J.~C., Fong, W., Kilpatrick, C.~D., Nicholl, M., \& Metzger, B.~D.
  2024.
\newblock \doarXiv{2409.02158}

\bibitem[{{Rastinejad} {et~al.}(2022){Rastinejad}, {Gompertz}, {Levan}, {Fong},
  {Nicholl}, {Lamb}, {Malesani}, {Nugent}, {Oates}, {Tanvir}, {de Ugarte
  Postigo}, {Kilpatrick}, {Moore}, {Metzger}, {Ravasio}, {Rossi}, {Schroeder},
  {Jencson}, {Sand}, {Smith}, {Ag{\"u}{\'\i} Fern{\'a}ndez}, {Berger},
  {Blanchard}, {Chornock}, {Cobb}, {De Pasquale}, {Fynbo}, {Izzo}, {Kann},
  {Laskar}, {Marini}, {Paterson}, {Escorial}, {Sears}, \&
  {Th{\"o}ne}}]{2022Natur.612..223R}
{Rastinejad}, J.~C., {Gompertz}, B.~P., {Levan}, A.~J., {et~al.} 2022, \nat,
  612, 223, \dodoi{10.1038/s41586-022-05390-w}

\bibitem[{{Saio} \& {Nomoto}(2004)}]{2004ApJ...615..444S}
{Saio}, H., \& {Nomoto}, K. 2004, \apj, 615, 444, \dodoi{10.1086/423976}

\bibitem[{{Sharon} \& {Kushnir}(2020)}]{2020ApJ...894..146S}
{Sharon}, A., \& {Kushnir}, D. 2020, \apj, 894, 146,
  \dodoi{10.3847/1538-4357/ab8a31}

\bibitem[{{Shen} \& {Bildsten}(2009)}]{2009ApJ...692..324S}
{Shen}, K.~J., \& {Bildsten}, L. 2009, \apj, 692, 324,
  \dodoi{10.1088/0004-637X/692/1/324}

\bibitem[{{Shen} {et~al.}(2009){Shen}, {Idan}, \&
  {Bildsten}}]{2009ApJ...705..693S}
{Shen}, K.~J., {Idan}, I., \& {Bildsten}, L. 2009, \apj, 705, 693,
  \dodoi{10.1088/0004-637X/705/1/693}

\bibitem[{{Siegel} {et~al.}(2019){Siegel}, {Barnes}, \&
  {Metzger}}]{2019Natur.569..241S}
{Siegel}, D.~M., {Barnes}, J., \& {Metzger}, B.~D. 2019, \nat, 569, 241,
  \dodoi{10.1038/s41586-019-1136-0}

\bibitem[{{Stergioulas} \& {Friedman}(1995)}]{1995ApJ...444..306S}
{Stergioulas}, N., \& {Friedman}, J.~L. 1995, \apj, 444, 306,
  \dodoi{10.1086/175605}

\bibitem[{{Suwa} {et~al.}(2007){Suwa}, {Takiwaki}, {Kotake}, \&
  {Sato}}]{2007PASJ...59..771S}
{Suwa}, Y., {Takiwaki}, T., {Kotake}, K., \& {Sato}, K. 2007, \pasj, 59, 771,
  \dodoi{10.1093/pasj/59.4.771}

\bibitem[{{Tanaka}(2016)}]{2016AdAst2016E...8T}
{Tanaka}, M. 2016, Advances in Astronomy, 2016, 634197,
  \dodoi{10.1155/2016/6341974}

\bibitem[{{Tanaka} \& {Hotokezaka}(2013)}]{2013ApJ...775..113T}
{Tanaka}, M., \& {Hotokezaka}, K. 2013, \apj, 775, 113,
  \dodoi{10.1088/0004-637X/775/2/113}

\bibitem[{{Troja} {et~al.}(2022){Troja}, {Fryer}, {O'Connor}, {Ryan},
  {Dichiara}, {Kumar}, {Ito}, {Gupta}, {Wollaeger}, {Norris}, {Kawai},
  {Butler}, {Aryan}, {Misra}, {Hosokawa}, {Murata}, {Niwano}, {Pandey},
  {Kutyrev}, {van Eerten}, {Chase}, {Hu}, {Caballero-Garcia}, \&
  {Castro-Tirado}}]{2022Natur.612..228T}
{Troja}, E., {Fryer}, C.~L., {O'Connor}, B., {et~al.} 2022, \nat, 612, 228,
  \dodoi{10.1038/s41586-022-05327-3}

\bibitem[{{Turk} {et~al.}(2011){Turk}, {Smith}, {Oishi}, {Skory}, {Skillman},
  {Abel}, \& {Norman}}]{2011ApJS..192....9T}
{Turk}, M.~J., {Smith}, B.~D., {Oishi}, J.~S., {et~al.} 2011, The Astrophysical
  Journal Supplement Series, 192, 9, \dodoi{10.1088/0067-0049/192/1/9}

\bibitem[{{Uenishi} {et~al.}(2003){Uenishi}, {Nomoto}, \&
  {Hachisu}}]{2003ApJ...595.1094U}
{Uenishi}, T., {Nomoto}, K., \& {Hachisu}, I. 2003, \apj, 595, 1094,
  \dodoi{10.1086/377248}

\bibitem[{{Varma} {et~al.}(2021){Varma}, {M{\"u}ller}, \&
  {Obergaulinger}}]{2021MNRAS.508.6033V}
{Varma}, V., {M{\"u}ller}, B., \& {Obergaulinger}, M. 2021, \mnras, 508, 6033,
  \dodoi{10.1093/mnras/stab2983}

\bibitem[{Virtanen {et~al.}(2020)Virtanen, Gommers, Oliphant, Haberland, Reddy,
  Cournapeau, Burovski, Peterson, Weckesser, Bright, {van der Walt}, Brett,
  Wilson, Millman, Mayorov, Nelson, Jones, Kern, Larson, Carey, Polat, Feng,
  Moore, {VanderPlas}, Laxalde, Perktold, Cimrman, Henriksen, Quintero, Harris,
  Archibald, Ribeiro, Pedregosa, {van Mulbregt}, \& {SciPy 1.0
  Contributors}}]{2020SciPy-NMeth}
Virtanen, P., Gommers, R., Oliphant, T.~E., {et~al.} 2020, Nature Methods, 17,
  261, \dodoi{10.1038/s41592-019-0686-2}

\bibitem[{{Wang} {et~al.}(2022){Wang}, {Liu}, \& {Chen}}]{2022MNRAS.510.6011W}
{Wang}, B., {Liu}, D., \& {Chen}, H. 2022, \mnras, 510, 6011,
  \dodoi{10.1093/mnras/stac114}

\bibitem[{{Waxman}(2017)}]{2017ApJ...842...34W}
{Waxman}, E. 2017, \apj, 842, 34, \dodoi{10.3847/1538-4357/aa713e}

\bibitem[{{W}es {M}c{K}inney(2010)}]{mckinney-proc-scipy-2010}
{W}es {M}c{K}inney. 2010, in {P}roceedings of the 9th {P}ython in {S}cience
  {C}onference, ed. {S}t\'efan van~der {W}alt \& {J}arrod {M}illman, 56 -- 61,
  \dodoi{10.25080/Majora-92bf1922-00a}

\bibitem[{{Woosley} \& {Baron}(1992)}]{1992ApJ...391..228W}
{Woosley}, S.~E., \& {Baron}, E. 1992, \apj, 391, 228, \dodoi{10.1086/171338}

\bibitem[{Wu {et~al.}(2022)Wu, Ricigliano, Kashyap, Perego, \&
  Radice}]{Wu:2021ibi}
Wu, Z., Ricigliano, G., Kashyap, R., Perego, A., \& Radice, D. 2022, Mon. Not.
  Roy. Astron. Soc., 512, 328, \dodoi{10.1093/mnras/stac399}

\bibitem[{{Yang} {et~al.}(2022){Yang}, {Ai}, {Zhang}, {Zhang}, {Liu}, {Wang},
  {Yang}, {Yin}, {Li}, \& {L{\"u}}}]{2022Natur.612..232Y}
{Yang}, J., {Ai}, S., {Zhang}, B.-B., {et~al.} 2022, \nat, 612, 232,
  \dodoi{10.1038/s41586-022-05403-8}

\bibitem[{{Yi} \& {Blackman}(1998)}]{1998ApJ...494L.163Y}
{Yi}, I., \& {Blackman}, E.~G. 1998, \apjl, 494, L163, \dodoi{10.1086/311192}

\bibitem[{{Yoon} \& {Langer}(2004)}]{2004A&A...419..623Y}
{Yoon}, S.~C., \& {Langer}, N. 2004, \aap, 419, 623,
  \dodoi{10.1051/0004-6361:20035822}

\bibitem[{{Yoon} \& {Langer}(2005)}]{2005A&A...435..967Y}
---. 2005, \aap, 435, 967, \dodoi{10.1051/0004-6361:20042542}

\bibitem[{{Yoon} {et~al.}(2007){Yoon}, {Podsiadlowski}, \&
  {Rosswog}}]{2007MNRAS.380..933Y}
{Yoon}, S.~C., {Podsiadlowski}, P., \& {Rosswog}, S. 2007, \mnras, 380, 933,
  \dodoi{10.1111/j.1365-2966.2007.12161.x}

\bibitem[{{Zhu} {et~al.}(2015){Zhu}, {Pakmor}, {van Kerkwijk}, \&
  {Chang}}]{2015ApJ...806L...1Z}
{Zhu}, C., {Pakmor}, R., {van Kerkwijk}, M.~H., \& {Chang}, P. 2015, \apjl,
  806, L1, \dodoi{10.1088/2041-8205/806/1/L1}

\end{thebibliography}
\bibliographystyle{aasjournal}



\end{document}